\documentclass[fleqn,10pt]{wlscirep}
\usepackage[utf8]{inputenc}
\usepackage{lineno}
\usepackage[T1]{fontenc}
\title{Conditional diffusion model  for inverse prediction \\ of process parameters and dendritic microstructures \\ from mechanical properties }

\author[1]{Arisa Ikeda}
\author[2]{Ryo Higuchi}
\author[2]{Tomohiro Yokozeki}
\author[3]{Katsuhiro Endo}
\author[1]{Yuta Kojima}
\author[1]{\\ Misato Suzuki}
\author[4,*]{Mayu Muramatsu}

\affil[1]{Graduate School of Science and Technology, Keio University, 3-14-1, Hiyoshi, Kohoku-ku, Yokohama, Kanagawa 223-8522, Japan}
\affil[2]{Department of Aeronautics and Astronautics, The University of Tokyo, 7-3-1, Hongo, Bunkyo-ku, Tokyo 113-8656, Japan}
\affil[3]{National Institute of Advanced Industrial Science and Technology, 1-1-1, Umezono, Tsukuba, Ibaraki 305-8568, Japan}
\affil[4]{Department of Mechanical Engineering, Keio University, 3-14-1, Hiyoshi, Kohoku-ku, Yokohama, Kanagawa 223-8522, Japan}

\affil[*]{muramatsu@mech.keio.ac.jp}

\begin{abstract}
In this study, we develop a conditional diffusion model that proposes the optimal process parameters and predicts the microstructure for the desired mechanical properties.
In materials development, it is costly to try many samples with different parameters in experiments and numerical simulations.
The use of data-driven inverse design method can reduce the cost of materials development.
This study develops an inverse analysis model that predicts process parameters and microstructures.
This method can be used for any material, but in this study it is applied to polymeric material, which is the matrix resin of carbon fiber reinforced thermoplastics as an example.
Matrix resins contain a mixture of dendrites, which are crystalline phases, and amorphous phases even after crystal growth is complete, and it is important to consider the microstructures consisting of the crystalline structure and the remaining amorphous phase to achieve the desired mechanical properties.
Typically, the temperature during forming affects the microstructures, which in turn affect the macroscopic mechanical properties.
The trained diffusion model can propose not only the processing temperature but also the microstructure when Young's modulus and Poisson's ratio are given.
The capability of our conditional diffusion model to represent complex dendrites is also noteworthy.
\end{abstract}

\begin{document}

\flushbottom
\maketitle

\thispagestyle{empty}

\section{Introduction}\label{sec1}

In materials development, experiments require trial and error with multiple parameters, which is costly. 
To solve this problem, numerical simulations have been studied in various fields. 
In order to accelerate the computation by numerical simulation, hardware performance-based approaches such as parallel computing using graphics processing units parallelism\cite{sakane2024mother} and software-based approaches using machine learning\cite{rezaei2024learning} have been used.
For example, machine learning is used to predict mechanical properties \cite{eidel2023deep, pathan2019predictions, maurizi2022predicting, minkowski2023predicting, onyelowe2025prediction}.
These precious numerical simulations with machine learning are limited to forward analyses.
Since forward analysis cannot infer the causes from effects, a data-driven approach to solve the inverse problem is required for more efficient material development. 
Hashemi et al. \cite{hashemi2021supervised} presented a supervised machine learning based computational methodology for the design of particulate multifunctional composite materials with desired thermal conductivity, using the design of particulate composites with liquid metal elastomer as a case study. 
The method consists of three phases: data generation in the first phase, discovery of complex relationship between the structure and properties using appropriate machine learning algorithms in the second phase, and inference of direct structure-property relationships and generation and visualization of candidate microstructures by inverse design framework in the third phase. 
This new supervised machine learning approach accelerates the prediction of the thermal conductivity of particle composites and enables the design of composites with desirable properties. 
Lee et al. \cite{lee2020dirty} aimed to provide a solution for inverse design even when data quality does not meet high standards, and presented a new design strategy employing two independent approaches: a metaheuristic-assisted inverse reading of conventional forward machine learning models and an atypical inverse machine learning model based on a modified variational autoencoder. 
They pinpointed several novel thermo-mechanically controlled processed steel alloy candidates, which were validated by a rule-based thermodynamic calculation tool. 
Bastek et al. \cite{bastek2023inverse} developed a method for the inverse design of nonlinear mechanical metamaterials using a video denoising diffusion model. 
The video denoising diffusion model is trained on full-field data of periodic stochastic cellular structures. 
They showed that the model can predict and tune the nonlinear deformation and stress response under compression in the large-strain regime.
Hiraide et al. \cite{hiraide2021application} developed a framework for forward analysis to predict Young's modulus from the phase-separated structure of polymer alloys and for inverse analysis to predict the structure from Young's modulus.
Forward analysis uses convolutional neural networks (CNNs).
For inverse analysis, random search is applied to the combined generative adversarial network (GAN) and CNN model.
Moreover, Hiraide et al. \cite{hiraide2023inverse} proposed a framework for designing material structures based on macroscopic properties. 
They use the results of the analysis of the two-dimensional phase separation structure of diblock copolymer melts as structural data. 
The stress data are obtained by the finite element analysis. 
The framework consists of a deep learning model that generates structures and a model that predicts the physical properties of the structures.
It generates structures with desired physical properties using random search.
Vlassis and Sun \cite{vlassis2023denoising} presented a denoising diffusion algorithm to discover microstructures with nonlinear fine-tuned properties using the open-source mechanical MNIST data set provided in Lejeune\cite{lejeune2020mechanical}.
They used a CNN architecture and a denoising diffusion algorithm, where the CNN architecture predicts the hyperelastic energy functional behavior under uniaxial extension, and the denoising diffusion algorithm generates targeted microstructures with desired constitutive responses.
In this way, several data-driven approaches for inverse analysis were proposed for materials development.

Carbon fiber reinforced thermosetting plastics (CFRTSs) and carbon fiber reinforced thermoplastics (CFRTPs) have excellent light weight and strength \cite{karatacs2018review}. 
These materials are increasingly applied to structural components of aircraft and automobiles \cite{hashish2013trimming,wan2021development}.
CFRTSs have the property of stiffening upon heating and not returning to its original state, whereas CFRTPs have the property of stiffening upon cooling and softening again upon heating. 
CFRTSs have been applied to many structures, but the material and molding costs are high.
In addition, it is difficult to recycle CFRTSs once stiffened, which poses a disposal problem \cite{chen2022circular, almushaikeh2023manufacturing}. 
On the other hand, CFRTPs have been attracting attention as a sustainable material due to its low molding cost and recyclability.
In CFRTPs, the mechanical properties depend not only on the carbon fibers, but also on the thermoplastic resin used as the base material.
The thermoplastic resins used in CFRTPs are crystalline resins, such as polyphenylene sulfide (PPS) and polyether ether ketone (PEEK), which have significant rigidity \cite{waddon1987crystal,nohara2006study}.
Crystalline and amorphous phases are mixed in thermoplastic resins, and it is important to consider the microstructures consisting of the crystalline structure and the remaining amorphous phase to achieve the desired mechanical properties.
As crystallization progresses, thermoplastic resins become stronger and stiffer.
This increases the overall strength and stiffness of the composite.
On the other hand, the amorphous phase contributes to the ductility of the composite.
Therefore, it is important to consider the balance of crystalline and amorphous phase and their arrangement.
In this study, we focus on PPS, a thermoplastic resin with low costs.
PPS has a mixture of crystalline and amorphous phases even when the crystals are fully grown.
It produces dendritic crystals called dendrites during solidification \cite{trivedi1994dendritic}.
Process parameters such as temperature during forming and cooling rate affect the microstructure, which in turn affects the macroscopic mechanical properties \cite{ye1995matrix,gao2000cooling, gao2001cooling2,gao2001cooling3,oshima2022high,oshima2023experimental}.
Thus, a method of predicting the mechanical properties of resins is required.

Higuchi et al. \cite{ryo2022multiphysics} and Takashima et al. \cite{takashima2024} developed a multiphysics analysis method for crystalline thermoplastic resins that links the forming conditions, the microstructure and the macroscopic mechanical properties.
Crystallization analysis was performed by using the phase-field method \cite{takaki2014phase,kobayashi1993modeling,bahloul2020enhanced}.
The homogenization analysis \cite{guedes1990preprocessing} of PPS, a crystalline thermoplastic resin, was conducted by using the extended finite element method (XFEM) \cite{higuchi2017numerical,higuchi2019evaluation,nagashima2016development}.
PPS was characterized to identify the parameters necessary for crystallization analysis.
The simulation results were validated by comparing them with experimental data.
The details of their studies \cite{ryo2022multiphysics,takashima2024} are as follows. 
They measured properties such as the glass transition temperature, crystallization temperature and melting point by differential scanning calorimetry (DSC). 
Moreover, they examined the relationship between the processing temperature and the mechanical properties by tensile tests. 
The phase-field model they developed reproduced various types of nucleation and polycrystal growth. 
Conventional methods for dendrite simulation divide the analysis domain into cells and discretely assign states to them, e.g., the cellular automata \cite{mf2001modified,beltran2003growth,beltran2004quantitative} and Monte Carlo \cite{plapp2000multiscale,lue2000volumetric} methods.
The other method represents the interface by connecting discrete points such as the front-tracking method \cite{juric1996front,zhu2007virtual}.
Higuchi et al. \cite{ryo2022multiphysics} used the phase-field method to describe the crystal growth of dendrites in thermoplastic resins, in terms of the interface movement between two different phases.
Then, they performed the homogenization analysis of the obtained microstructures using XFEM to determine the mechanical properties such as Young's modulus. 
By comparing the analytical results with the experimental results, it is shown that their proposed method has a certain validity.
Their method is limited to forward analysis, and it is not possible to predict the processing temperature or microstructure from the mechanical properties.

Thus, methods of predicting mechanical properties from microstructures and  microstructures from mechanical properties have been developed.
However, there is no scheme for proposing process parameters and material microstructures related to how materials should be made. 
If process parameters such as processing temperature and cooling rate are proposed to produce materials with the desired mechanical property, it will be possible to develop materials efficiently by eliminating the need for trial-and-error experiments using different parameters.

\begin{figure}[bt]
    \centering
    \includegraphics[width=0.9\columnwidth]{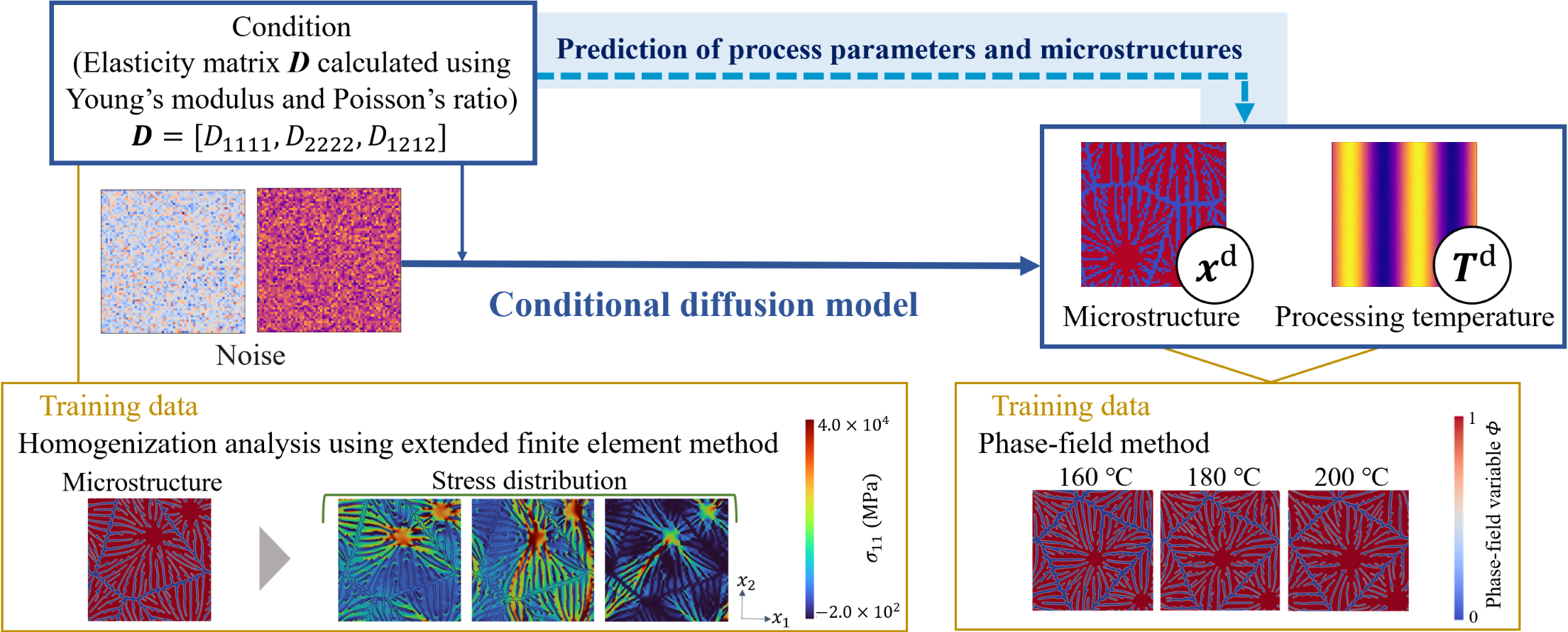}
    \caption{Diagram of conditional diffusion model}
    \label{fig:conditionaldiffusion_model}
\end{figure}

The purpose of this study is to develop a conditional diffusion model that proposes the optimal process parameters and predicts the microstructure on the basis of the desired mechanical properties. 
As shown in Fig. \ref{fig:conditionaldiffusion_model}, the inverse analysis model is applied to thermoplastic resins in this study to suggest the process temperature and predict the microstructure for the desired Young's modulus and Poisson's ratio.

This paper is organized as follows. 
The details of the methods for crystal growth, homogenization analysis using XFEM and conditional diffusion model are described in Section 2. The results and discussion of this study are given in Section 3. Finally, the conclusions are presented in Section 4.

\section{Methods}\label{sec2}
\subsection{Crystal growth}
This study introduces two models for the crystal nucleation and growth.
The models are used to generate the data of microstructures depending on the temperature.
In this study, the crystallization and ambient temperatures are the same owing to the assumption of isothermal forming.
Therefore, we consider the crystallization temperature as a parameter that can be specified in the process.

\subsubsection{Crystal growth by phase-field method}
The phase-field method is a method of describing the movement of surfaces by solving the time evolution equation of the phase-field variable $\phi$.
In this method, the dimensionless variable $\phi$, which varies between $0$ and $1$ where $\phi=1$ for the crystalline phase and $\phi=0$ for the amorphous phase, is used to show the spatial distribution of the crystalline and amorphous phases.

The crystal growth is described by the Allen--Cahn and heat conduction equations respectively as follows \cite{bahloul2020enhanced}:
\begin{equation}
\label{Allen-Cahn}
\frac{\partial \phi}{\partial t^\text p}=-M^\text p \frac{\delta F^\text p(\phi)}{\delta \phi},
\end{equation}
\begin{equation}
\label{heat_equilibrium}
\frac{\partial T^\text p}{\partial t^\text p}=\alpha\nabla^2T^\text p+\frac{\Delta H}{C_\text p}\frac{\partial \phi}{\partial t^\text p},
\end{equation}
where $\phi$ is the phase-field variable, $t^\text p$ is the time, $M^\text p$ is the phase-field mobility, $F^\text p$ is the free energy of the system, $T^\text p$ is the field temperature, $\alpha$ is the thermal diffusivity, $\Delta H$ is the latent heat and $C_\text p$ is the specific heat at a constant pressure.
The total free energy of the system is expressed as follows \cite{kobayashi1993modeling}:
\begin{equation}
\label{total energy}
F^\text p(\phi)=\int\left[f^\text p_{\text {doub }}(\phi)+f^\text p_{\text {grad }}(\phi)\right] \mathrm{d} V,
\end{equation}
where $f^\text p_\text {doub}$ is the double-well potential and $f^\text p_\text {grad}$ is the gradient energy density. 
Here, $f^\text p_\text {doub}$, $f^\text p_\text {grad}$, $m^\text p(\hat{T}^\text p)$ and $\hat{T}^\text p$ are respectively expressed as
\begin{equation}
\label{double potential}
f^\text p_{\text {doub }}(\phi)=W \int_0^\phi \phi\left(\frac{1}{2}-\phi-m^\text p(\hat{T}^\text p)\right)(1-\phi) \mathrm{d}\phi, 
\end{equation}
\begin{equation}
\label{grad energy}
f^\text p_{\text {grad  }}(\phi)=\frac{1}{2}\epsilon^2(\nabla\phi)^2,
\end{equation}
\begin{equation}
\label{m(T)}
m^\text p(\hat{T}^\text p)=\frac{a_\text k}{\pi} \arctan (\gamma(1-\hat{T}^\text p)),
\end{equation}
\begin{equation}
\label{T}
\hat{T}^\text p=\frac{T^\text p-T^\text p_\text c}{T^\text p_\text m-T^\text p_\text c},
\end{equation}
where $W$ is the height of the energy barrier, $\hat{T}^\text p$ is the dimensionless temperature and $\epsilon$ is the coefficient of interface energy gradient.
The constants are set as $a_\text k=0.9$ and $\gamma=10$. 
From Eq. (\ref{T}), we find that the crystal growth depends on the crystallization temperature $T^\text p_\text c$.

\subsubsection{Formation of crystal nuclei}
For the modeling of crystal nucleation, in this study, we introduce the model proposed by Pantani et al. \cite{pantani2005modeling} as
\begin{equation}
\label{crystal nuclei}
\frac{d N(T^\text p(t^\text p))}{d t^\text p}=N_0 \exp \left[-\frac{C_1}{\left(T^\text p(t^\text p)-T^\text p_{\infty}\right)}\right] \exp \left[-\frac{C_2\left(T^\text p(t^\text p)+T^\text p_\text m\right)}{T^\text p(t^\text p)^2\left(T^\text p_\text m-T^\text p(t^\text p)\right)}\right],
\end{equation}
where $\frac{d N(T^\text p(t^\text p))}{d t^\text p}$ is the nucleation rate, $N$ is the nucleation density, $T^\text p$ is the field temperature, $t^\text p$ is the time, $N_0$ is a constant independent of temperature, $T^\text p_{\infty}$ is the temperature at which molecular motion stops completely, $C_1$ and $C_2$ are nucleation rate parameters and $T^\text p_\text m$ is the melting point obtained at a particular crystallization temperature $T^\text p_\text c$.
The Hoffman--Weeks plot shows the crystallization temperature $T^\text p_\text c$ on the horizontal axis and the melting point $T^\text p_\text m$ on the vertical axis.
DSC \cite{schick2009differential} gives the melting point from the starting temperature of the melting peak of the DSC curve. 
Thus, nucleation depends on the crystallization temperature $T^\text p_\text c$.
In Eq. (\ref{crystal nuclei}), ${T^\text p_\text m}=a{T^\text p_\text c}+b$, where $T^\text p_\text c$ is the crystallization temperature (ambient temperature), $a = 0.0948$ is the slope of the approximate line between the crystallization temperature and the melting point obtained from the Hoffman--Weeks plot, and $b = 253.7$ is the intercept of the line \cite{kato2024experimental}.
Although this model assumes homogeneous nucleation in which the nucleation occurs stochastically on the basis of nucleation rate, the actual nucleation is heterogeneous because, for example, unknown inclusions can accelerate the nucleation.
To consider this effect, a few initial nuclei are initially introduced in this study.

\subsubsection{Generation of crystal microstructure by the phase-field method}
To generate the microstructure of thermoplastic resins for machine learning, we use the phase-field method.
The size of a nucleus is equal to one grid for simplicity.
The analysis is performed at three crystallization temperatures: 160 \, ${}^\circ$C, 180 \, ${}^\circ$C and 200 \, ${}^\circ$C.
Table \ref{PF_condition} shows the conditions used in the analysis by the phase-field method.
The condition parameters in the phase-field method are obtained from the references \cite{bahloul2020enhanced, lovinger1985kinetic}.

\begin{table}[tb]
    \centering
    \caption{
        {Analysis conditions for phase-field method}{\textmd{}}
    }
    \label{PF_condition}
    \begin{tabular}{c c}\hline
        Number of calculation steps & 20,000\\
        Number of grid points &$320\times320$\\
        Crystallization temperature $T^{\text p}_{\text c}$ & 160 \, ${}^\circ$C, 180 \, ${}^\circ$C, 200 \, ${}^\circ$C \\
        Temperature at maximum crystallization rate $T^\text p_{\text c \, \rm{max}}$ & 180 \, ${}^\circ$C \\
        Glass transition temperature $T^\text p_\text g$ & 100 \, ${}^\circ$C \\
        Number of initial nuclei & 2 \\
        Boundary condition & Periodic boundary condition \\ \hline
    \end{tabular}
\end{table}

\subsection{Homogenization analysis using XFEM}
\subsubsection{XFEM}
The elasticity matrix $\boldsymbol{D}$ used as the condition of the diffusion model is obtained by homogenization analysis using XFEM for thermoplastic resins generated by the phase-field method.
At the interface between the crystalline and amorphous phases, the strain is discontinuous and the local modification of the interpolation function allows accurate approximation.
The nodes for XFEM are generated on the basis of the differentially discretized grid for the phase-field method and the triangular elements are generated by dividing the square grid into two parts.
In addition, subelements are introduced for elements containing interfaces to integrate discontinuous functions with high accuracy. 
In XFEM, the displacement is expressed as
\begin{equation}
\label{X_displacement}
\boldsymbol{u}^{\mathrm{h}}=\sum_{I=1} N^\text x_I(\boldsymbol{x}^\text x) (\boldsymbol{u}_I+R(\boldsymbol{x}^\text x) \boldsymbol{a}^\text x_I),
\end{equation} 
where $I$ is the node, $N^\text x_I(\boldsymbol{x}^\text x)$ is the shape functions, $\boldsymbol{x}^\text x$ is the coordinates, $\boldsymbol{u}_I$ is the vector of node displacement, $R(\boldsymbol{x}^\text x)$ is the enriched function to be introduced locally in the interpolating function and $\boldsymbol{a}^\text x_I$ is the nodal degree of freedom for the basis function $N^\text x_I(\boldsymbol{x}^\text x)R(\boldsymbol{x}^\text x)$ added by the enriched function. 
The ramp function proposed by Mo{\"e}s et al. \cite{moes2003computational}, which is introduced as an enriched function, can express the continuity of displacement and its derivative, which is the discontinuity of strain, and is expressed as
\begin{equation}
R(\boldsymbol{x}^\text x)=\sum_{I=1} |\psi_I|N^\text x_I(\boldsymbol{x}^\text x)-\left|\sum_{I=1} \psi_I N^\text x_I(\boldsymbol{x}^\text x)\right|,
\end{equation}
where $\psi_I$ is the level set function at node $I$ and is expressed in terms of the phase-field variable $\phi$ as
\begin{equation}
\psi=2\phi-1.
\end{equation}
Thus, $\psi=0$, which means that the phase-field variable $\phi=0.5$, is recognized as the interface between the crystalline and amorphous phases.

\subsubsection{Elasticity matrix calculated by homogenization analysis using XFEM}
In this study, we use microstructure analysis tools that combine XFEM and the homogenization method \cite{higuchi2019evaluation}.
We obtain the stress distribution using XFEM. 
Then, we obtain the mean stress by the homogenization method assuming a homogeneous body.
The key-degree-of-freedom method proposed by Li et al. \cite{li2011unit}, which imposes boundary conditions via the additional degree of freedom apart from the model mesh, makes it easier to treat macroscopic stresses and strains. 
In this study, we use the nonlinear XFEM code combined with phase-field simulation \cite{nagashima2016development}.

\begin{figure}[t]
    \centering
    \includegraphics[width=0.3\columnwidth]{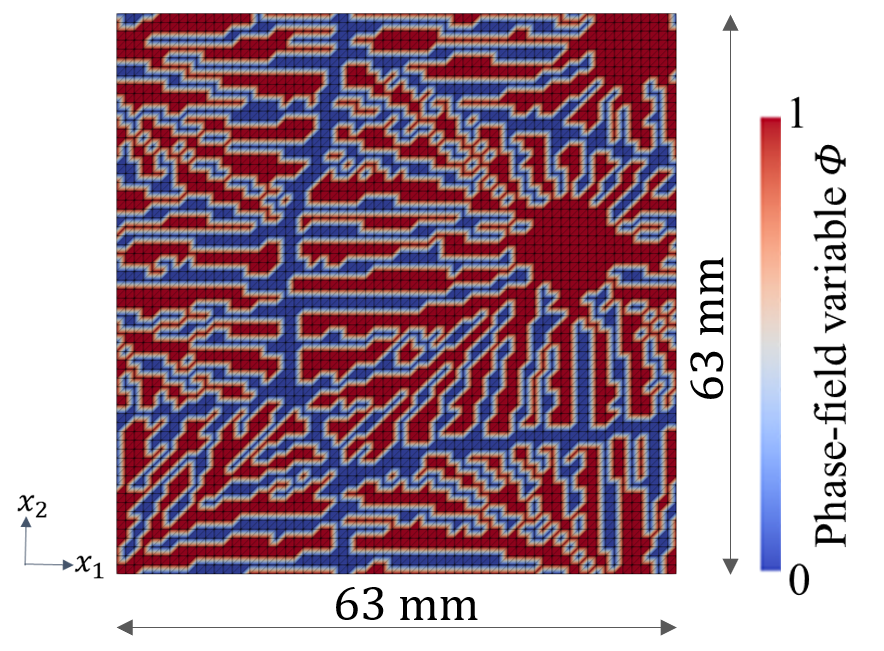}
    \caption{Analysis size of XFEM}
    \label{fig:XFEM_size}
\end{figure}

Table \ref{XFEM_condition} shows the conditions for XFEM.
The analysis size of XFEM is shown in Fig. \ref{fig:XFEM_size}.
The length order of this phase-field method is nondimensionalized.
Since dimensions are not relevant in homogenization analysis using XFEM, the dimensionless model is directly loaded into a program with an$\ \mathrm{mm\mathchar`-t\mathchar`-s}$ unit system and solved on the $\mathrm{mm}$ order.
Table \ref{XFEM_material} shows the physical properties of the PPS used for XFEM.
The elasticity matrix $\boldsymbol{D}$ is obtained by the identified Young’s modulus and Poisson’s ratio \cite{ryo2022multiphysics,takashima2024} (\ref{secA1}).

\begin{table}[tb]
    \centering
    \caption{
        {Analysis conditions for XFEM}{\textmd{}}
    }
    \label{XFEM_condition}
    \begin{tabular}{c c}\hline
        Size of analysis area &$63{\ \mathrm{mm}}\times63{\ \mathrm{mm}}$\\
        Incremental time  & 1.0 s\\
        Newton--Raphson method convergence index & 0.1 \\
        Boundary condition & Periodic boundary condition \\ \hline
    \end{tabular}
\end{table}
\begin{table}[tb]
  \caption{Physical properties of crystalline and amorphous phases in homogenization analysis using XFEM}{\textmd{}}
  \label{XFEM_material}
  \centering
  \begin{tabular}{ccc}
    \hline
    & Crystal & Amorphous  \\
    \hline \hline
    Young's modulus $E$  & 28,000 MPa \cite{nishino1992elastic}  & 150 MPa \cite{ryo2022multiphysics,takashima2024} \\
    Poisson's ratio $\nu$ & 0.2 \cite{ryo2022multiphysics,takashima2024} & 0.4 \cite{ryo2022multiphysics,takashima2024} \\
    \hline
  \end{tabular}
\end{table}

\subsection{Conditional diffusion model}
\begin{figure}[tb]
    \centering
    \includegraphics[width=0.8\columnwidth]{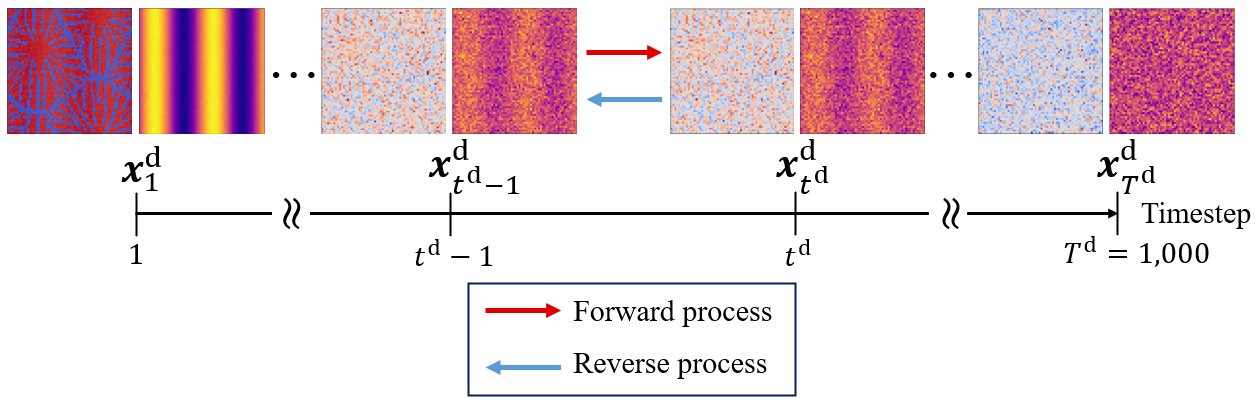}
    \caption{Forward and reverse processes of conditional diffusion model}
    \label{fig:diffusionnoise}
\end{figure}

\begin{figure}[tb]
    \centering
    \includegraphics[width=0.75\columnwidth]{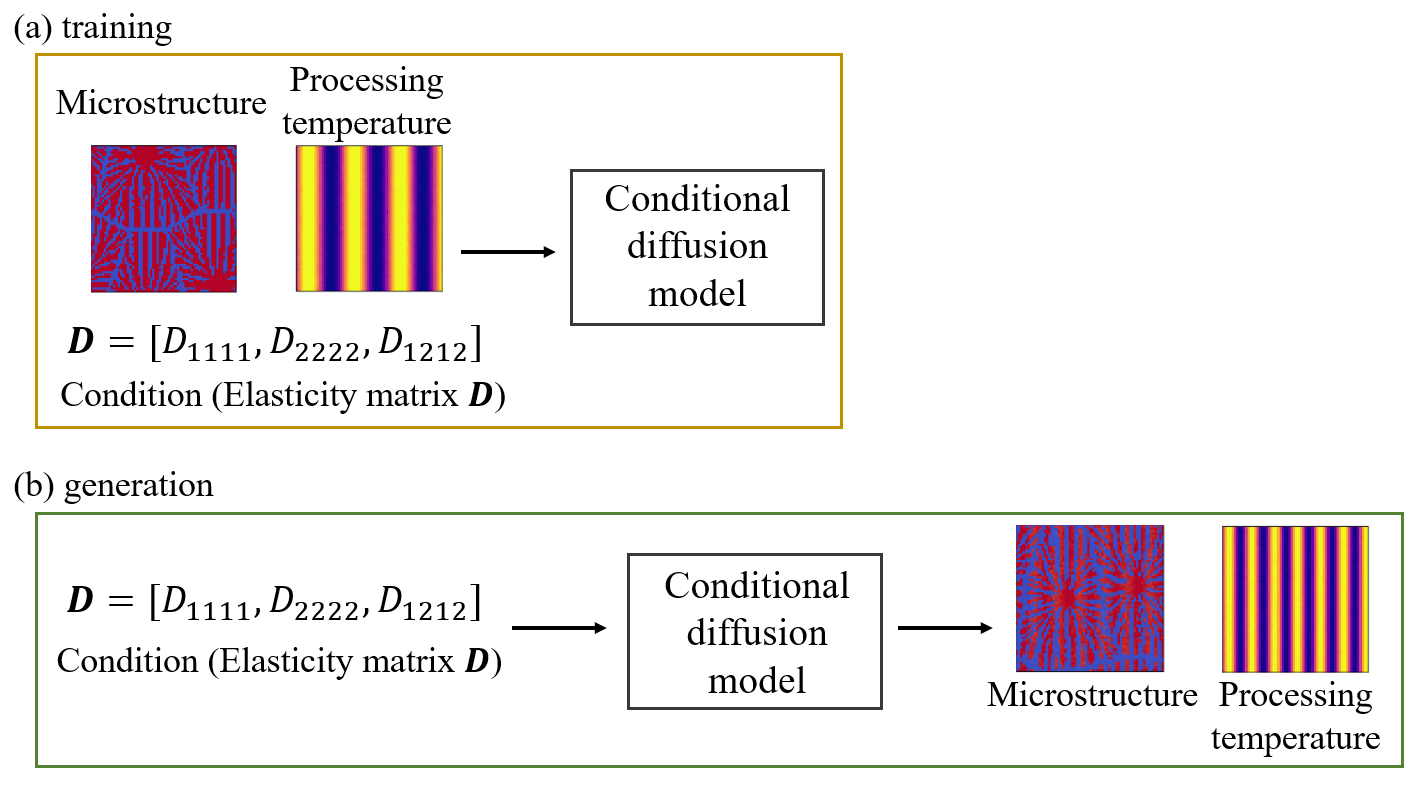}
    \caption{Diagram of (a) training and (b) generation of conditional diffusion model}
    \label{fig:trainingandgeneration}
\end{figure}

In this study, we employ a diffusion model for predicting the process parameters and the microstructures for the desired mechanical properties.
The diffusion model \cite{sohl2015deep,denoising,yang2023diffusion, dhariwal2021diffusion,ho2022classifier,croitoru2023diffusion} was proposed by Sohl-Dickstein et al. \cite{sohl2015deep} and improved as the denoising diffusion probabilistic model (DDPM) by Ho et al. \cite{denoising}.
The model used in this study is DDPM with condition, which is the called conditional diffusion model.
The conditional diffusion model consists of the forward and reverse processes, as shown in Fig. \ref{fig:diffusionnoise}.
In the forward process, Gaussian noise is added to the images at each timestep using Markov chains.
The forward process is defined as follows \cite{denoising}:
\begin{equation}
\label{q_theta}
q(\boldsymbol{x}_{1:T^\text d}^\text d|\boldsymbol{x}_0^\text d) = \prod_{t^\text d=1}^{T^\text d} q\left(\boldsymbol{x}^\text d_{t^\text d}|\boldsymbol{x}^\text d_{t^\text d-1}\right),
\end{equation}
\begin{equation}
\label{q}
q(\boldsymbol{x}_{t^\text d}^\text d|\boldsymbol{x}^\text d_{t^\text d-1}) = \mathcal{N}\left(\boldsymbol{x}^\text d_{t^\text d};\sqrt{1-\beta_{t^\text d}}\boldsymbol{x}^\text d_{t^\text d-1},\beta_{t^\text d}\boldsymbol{I}\right),
\end{equation}
where $\boldsymbol{x}^\text d_0$ is the data, $\boldsymbol{x}^\text d_{T^\text d}$ is the noise, $\boldsymbol{x}^\text d_1\text {--}\boldsymbol{x}^\text d_{T^\text d}$ are the latents, $t^\text d$ is the arbitrary timestep, $T^\text d$ is the total number of timesteps for noise addition and removal, $q$ is the stochastic process of the forward process, $\mathcal{N}$ is the normal distribution, $\beta_1, \ldots ,\beta_{T^\text d}$ are the variance schedule, $\sqrt{1-\beta_{t^\text d}}\boldsymbol{x}^\text d_{t^\text d-1}$ is the mean and $\beta_{t^\text d}\boldsymbol{I}$ is the variance.
In the reverse process, U-Net is used to remove the noise at each timestep.
During training, the model learns parameters for generating images from noise by repeatedly adding to and removing noise from images in the forward and reverse processes.
The reverse process is defined as follows \cite{denoising}:
\begin{equation}
\label{p_theta}
p_\theta(\boldsymbol{x}^\text d_{0:{T^\text d}}) = p(\boldsymbol{x}^\text d_{T^\text d}) \prod_{t^\text d=1}^{T^\text d} p_\theta(\boldsymbol{x}^\text d_{t^\text d-1}|\boldsymbol{x}^\text d_{t^\text d}),
\end{equation}
\begin{equation}
\label{p}
p_\theta(\boldsymbol{x}^\text d_{t^\text d-1}|\boldsymbol{x}^\text d_{t^\text d}) = \mathcal{N}\left(\boldsymbol{x}^\text d_{t^\text d-1};\boldsymbol{\mu}_\theta (\boldsymbol{x}^\text d_{t^\text d},t^\text d),\sum\nolimits_{\theta}(\boldsymbol{x}^\text d_{t^\text d},t^\text d)\right),
\end{equation}
where $p$ is the stochastic process of the reverse process, $p_\theta$ is the stochastic process of the reverse process using the neural network and $\boldsymbol{\mu}_\theta$ is parameterized as a neural network.
As shown in Fig. \ref{fig:trainingandgeneration}, the training data are the microstructures of thermoplastic resins generated by the phase-field method, the processing temperatures specified in the phase-field analysis and the elasticity matrix $\boldsymbol{D}$ obtained by homogenization analysis using XFEM for the microstructure generated by the phase-field analysis.
For the training data of processing temperature, we generate image patterns of processing temperature (IPPT).
Note that IPPT does not mean the distribution of temperatures, but gives a pattern for temperature conditions.

\begin{figure}[tb]
    \centering
    \includegraphics[width=0.9\columnwidth]{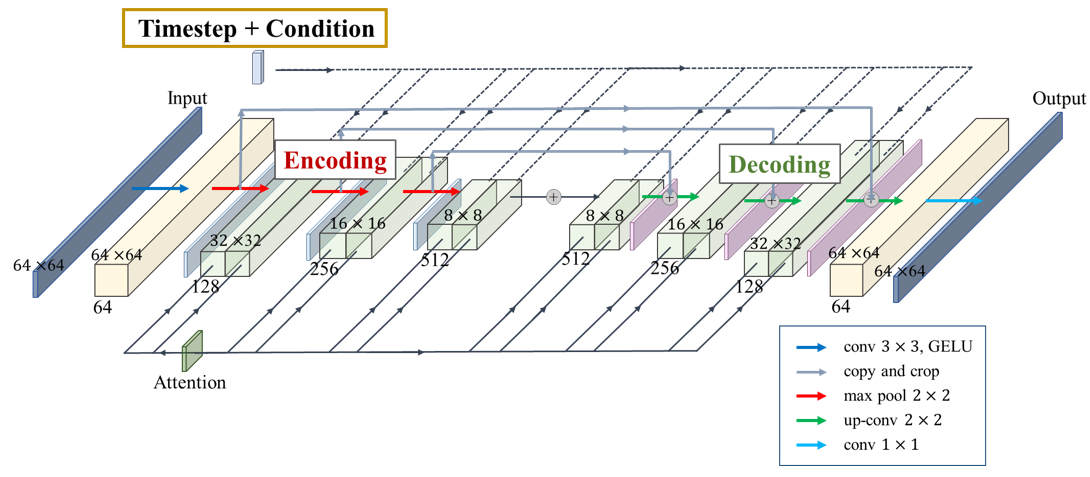}
    \caption{Structure of ST-UNet \cite{huang2023noise2music}}
    \label{fig:UNet}
\end{figure}

Noise is added to and removed from images of the microstructures and IPPTs.
In the process of image generation, only the reverse process of the diffusion model is used to generate images from noise distributions.
We use U-Net to remove the noise at each timestep.
The U-Net used in this study is the spatiotemporal U-Net (ST-UNet) \cite{yu2019st}, which includes the time information of noise-adding and noise-removing steps, as shown in Fig. \ref{fig:UNet} \cite{huang2023noise2music}.
U-Net is a method developed by Ronneberger et al. \cite{ronneberger2015u} for semantic segmentation, in which features are extracted from original images by an encoder and the obtained images are reconstructed in the same size as the original images by a decoder on the basis of the extracted features. 
Detailed information such as location information is not lost by maintaining feature maps with skip connections at each process.
It is possible to change the method of removing for each condition by considering the information of the timestep in the processes of adding and removing noise.
In this study, the component of the elasticity matrix $\boldsymbol{D}$ is used as the condition.
In U-Net, the condition and time information are added to the image at each of the encoding and decoding processes, as shown in Fig. \ref{fig:UNet}.
Then, the condition and time information are convolved with the image.

The training data are the microstructures of thermoplastic resins, IPPTs and the elasticity matrix $\boldsymbol{D}$.
The microstructures are generated by the phase-field method, IPPT means the temperature specified in the phase-field analysis and the elasticity matrix $\boldsymbol{D}$ is obtained by homogenization analysis using XFEM for the microstructures generated by the phase-field method.
We employ images of microstructures upon crystal growth completion.
Images of the microstructures of dendrites are compressed for use in training machine learning.
The mean values of $5 {\ \mathrm{pixel}}\times5 {\ \mathrm{pixel}}$ regions are calculated for each $320 {\ \mathrm{pixel}}\times320 {\ \mathrm{pixel}}$ analysis grid to be compressed to $64 {\ \mathrm{pixels}}\times64 {\ \mathrm{pixels}}$.
In addition, continuous phase-field variables are binarized during compression.
Moreover, we perform data augmentation on the microstructures generated by the phase-field method.
The $180^\circ$ rotation, inversion with respect to the $x_1$ axis, inversion with respect to the $x_1$ axis and $180^\circ$ rotation are conducted for the same elasticity matrix $\boldsymbol{D}$ value.

\begin{figure}[tb]
    \centering
    \includegraphics[width=0.35\columnwidth]{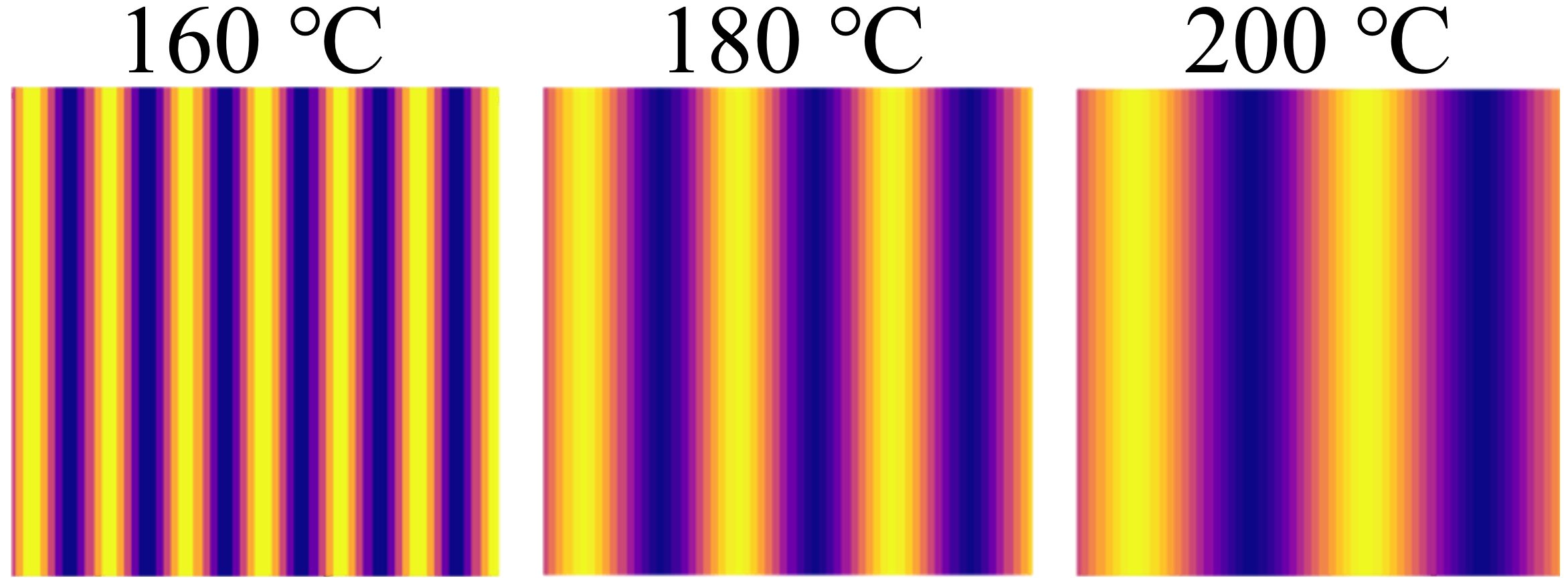}
    \caption{Image patterns of processing temperature (IPPT)}
    \label{fig:3tempr_img}
\end{figure}

To classify the temperature on the basis of the images, we create different patterns of images of IPPT at each temperature.
We specify $f=\frac{2.0}{T^\text p_\text c - 140} \, \rm{Hz}$ so that the frequency is a function of the crystallization temperature and $L^\text x(x)=A\sin({2\pi fx})$ to produce the sinusoidal stripe image as IPPT shown in Fig. \ref{fig:3tempr_img}, where $L^\text x$ is the luminance, $A$ is the amplitude, $f$ is the frequency, $x$ is the abscissa coordinate and $T^\text p_\text c$ is the crystallization temperature.
The reason for using sinusoidal stripe images as the training data is that these images can be used for determining the temperature through Fourier transform using the determined frequency.
The data of microstructures and IPPT are $64 {\ \mathrm{pixels}}\times64 {\ \mathrm{pixels}}\times1 {\ \mathrm{channel}}$, for a total of $2 {\ \mathrm{channels}}$.

The elasticity matrix $\boldsymbol{D}$ given as the condition is a vector with three components.
The conditions used in training are $D_{1111}$, $D_{2222}$ and $D_{1212}$, which are normalized from $0$ to $1$.
For the diffusion model, the time information during noise addition and removal is important.
The model is trained by adding 256-dimensionally expanded conditions to 256-dimensionally expanded time information.
Specifically, the original three-component condition values are repeated to extend the dimension, and the extra components are filled with zeros.
The time information is extended to 256 dimensions by positional encoding.
By adding the condition to the time information, we can change the weight of each condition so that the diffusion model learns a different way to remove noise depending on the condition.

\begin{figure}[tb]
    \centering
    \includegraphics[width=0.75\columnwidth]{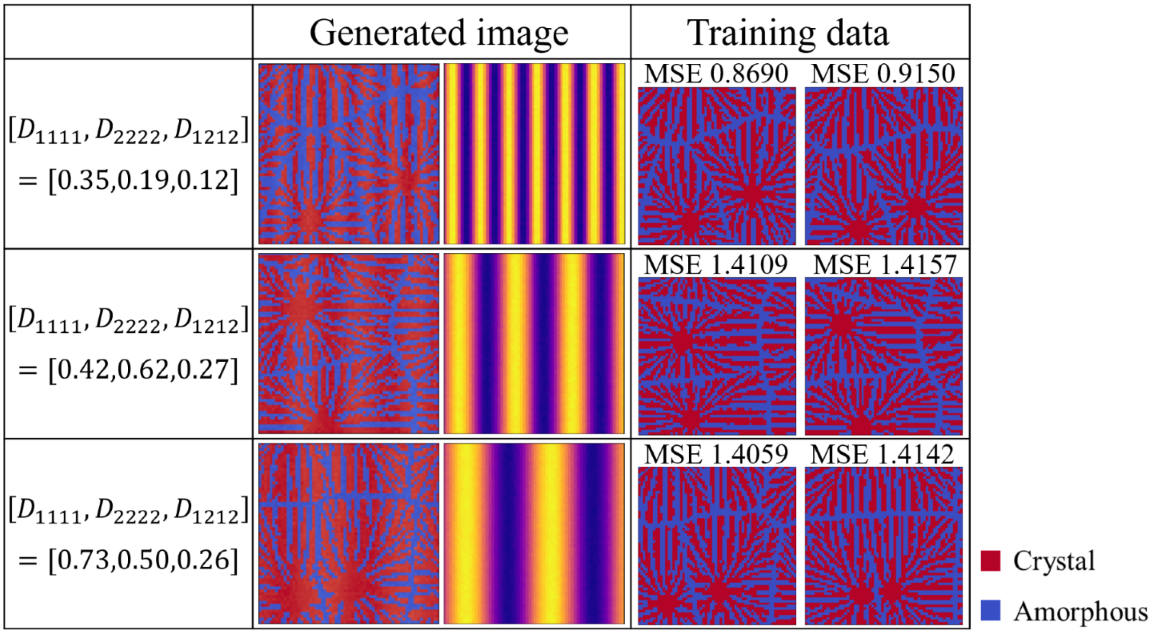}
    \caption{Comparing MSEs of training data and generated data to examine the model generalization performance}
    \label{fig:MSE_testcondition}
\end{figure}

Table \ref{trainingdata} shows the number and size of data used for training.
To avoid overfitting, we stopped training before an upward trend in the loss of the validation data.
By comparing the mean squared errors (MSEs) of the images generated using the trained model and the training data, we confirmed the generalization performance of the model. 
The value of MSE should be large because we want to generate data that is different from the training data.
As shown in Fig. \ref{fig:MSE_testcondition}, MSE is sufficiently large to show that the generated microstructure is different from the microstructure used for training.

\begin{table}[t]
    \centering
    \caption{Data used to train \text{$64{\ \mathrm{pixels}}\times64{\ \mathrm{pixels}}$}}
    \label{trainingdata}
    \begin{tabular}{c c}\hline
        Number of training data & 13,908\\
        Number of validation data & 1,728\\
        Number of test data & 432\\
        Image size &  $64{\ \mathrm{pixels}}\times64{\ \mathrm{pixels}}$\\
        Minibatch size & 15 \\ 
        Number of epochs & 1,189 \\ \hline
    \end{tabular}
\end{table}

\clearpage

\section{Results and Discussion}
\label{sec3}
\subsection{Dataset}
\begin{figure}[b]
    \centering
    \includegraphics[width=0.7\columnwidth]{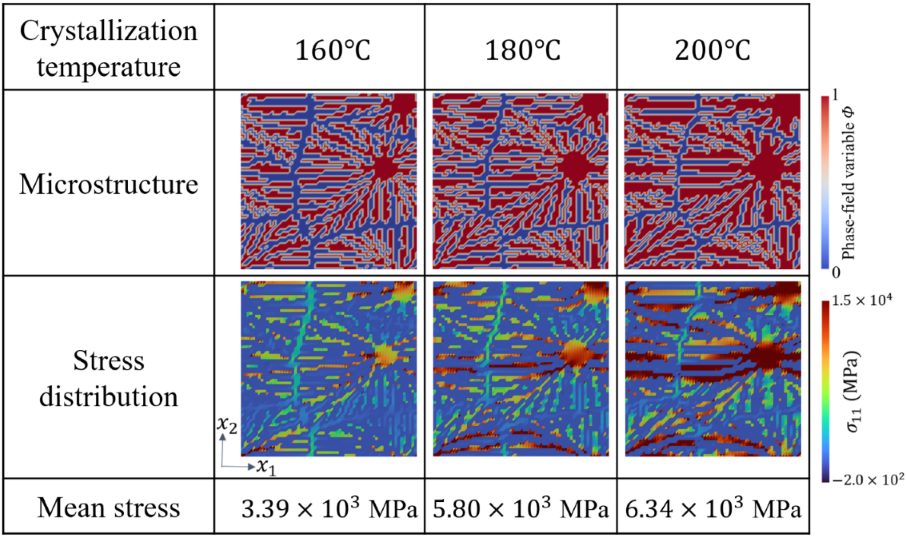}
    \caption{Examples of training data of the microstructures, the distribution of the $x$-directional tensile stress $\sigma_{11}$ and the mean stress}
    \label{fig:pf_XFEM_result}
\end{figure}

Fig. \ref{fig:pf_XFEM_result} shows examples of training data of the microstructures generated by the phase-field method, the distribution of the $x$-directional tensile stress $\sigma_{11}$ and the mean stress obtained by homogenization analysis using XFEM (under the condition of Eq. (\ref{sig11})).
The microstructures indicate that the higher the crystallization temperature, the thicker the crystal chains grow.
The stress distribution shows that high stresses are generated near the nuclei toward the direction of the imposed strain.
Moreover, increasing the crystallization temperature increases the values of stress distribution and mean stress.

\subsection{Validation of proposed processing temperatures and microstructures}
\begin{figure}[tb]
    \centering
    \includegraphics[width=0.75\columnwidth]{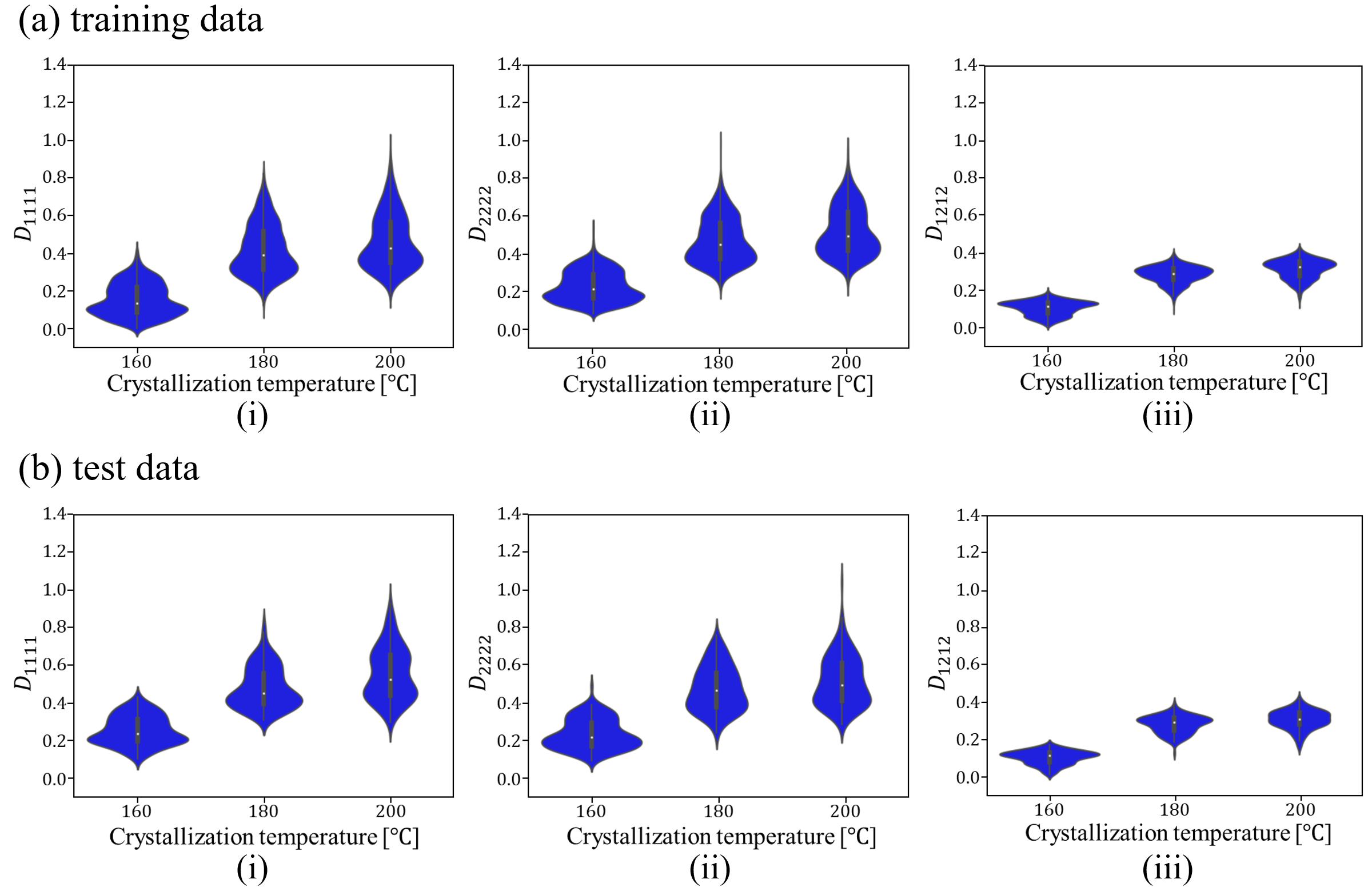}
    \caption{Relationship between crystallization temperature and elasticity matrix (i) $D_{1111}$, (ii) $D_{2222}$ and (iii) $D_{1212}$ in (a) training and (b) test data}
    \label{fig:trainandtest_violin}
\end{figure}

\begin{figure}[tb]
    \centering
    \includegraphics[width=0.75\columnwidth]{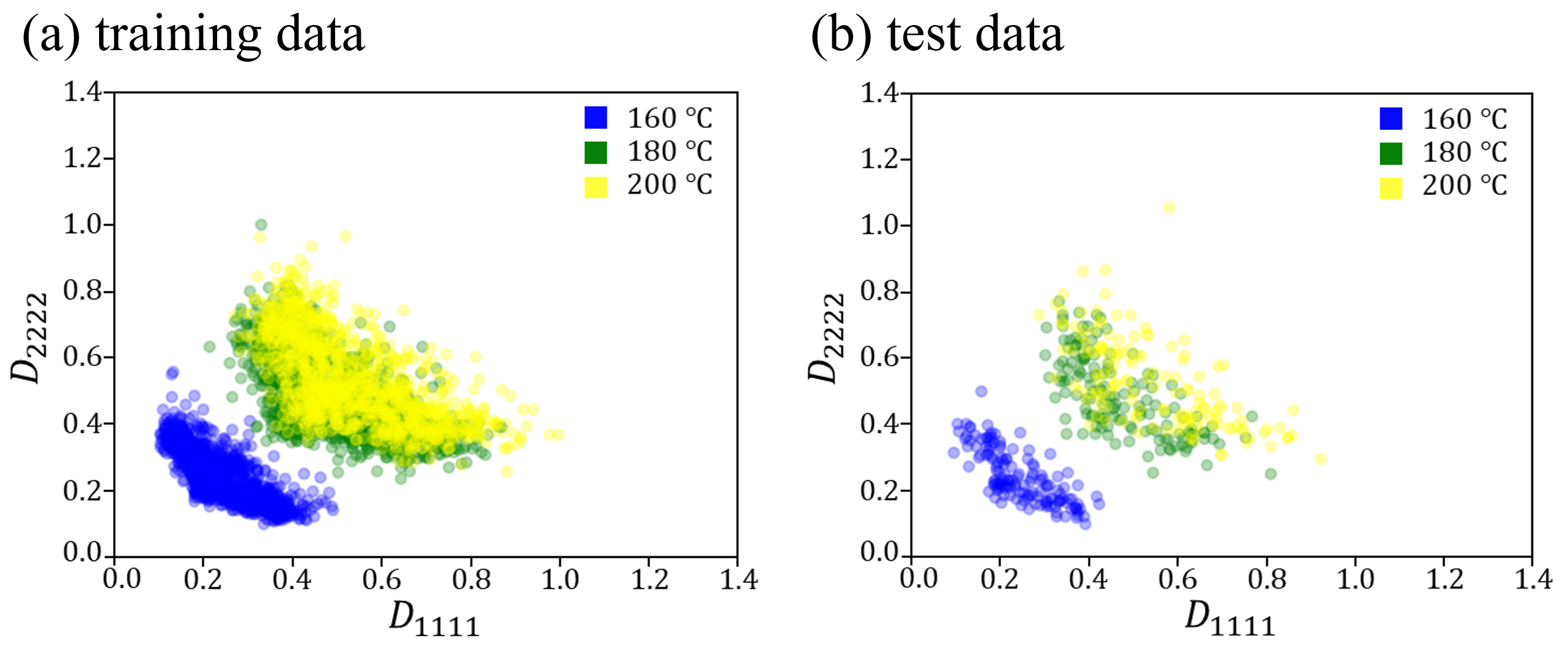}
    \caption{Relationship between $D_{1111}$ and $D_{2222}$ for each crystallization temperature in (a) training and (b) test data}
    \label{fig:trainandtest_plot}
\end{figure}

Figs. \ref{fig:trainandtest_violin} and \ref{fig:trainandtest_plot} show the relationship between the crystallization temperature specified in the phase-field analysis and the elasticity matrix $\boldsymbol{D}$ obtained by XFEM for the microstructures generated by the phase-field method for (a) training  and (b) test data, respectively.
The violin plots in Fig. \ref{fig:trainandtest_violin} show the distribution of the data with the horizontal axis for the crystallization temperature and the vertical axis for the value of each component of the elasticity matrix $\boldsymbol{D}$.
Each violin plot shows that there is a large amount of data in the bulging part of the plot and a large scatter of data in the vertically long part of the plot.
The black rectangle in the plot represents the interquartile range of the box plot, and the white dot at the center represents the median value.
Fig. \ref{fig:trainandtest_plot} shows scatter plots of $D_{1111}$ on the horizontal axis and $D_{2222}$ on the vertical axis, with different colors indicating the crystallization temperature. 
These plots show that the higher the crystallization temperature, the larger the value of each component of the elasticity matrix $\boldsymbol{D}$.

\begin{figure}[tb]
    \centering
    \includegraphics[width=0.8\columnwidth]{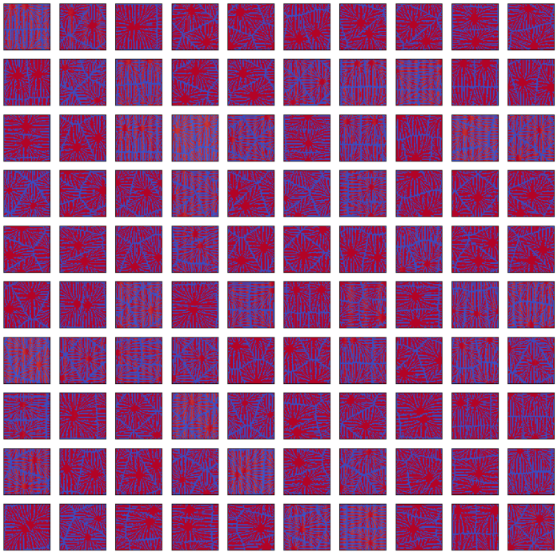}
    \caption{Predicted microstructures generated under conditions of test data}
    \label{fig:testdata_condition}
\end{figure}

\begin{figure}[tb]
    \centering
    \includegraphics[width=0.8\columnwidth]{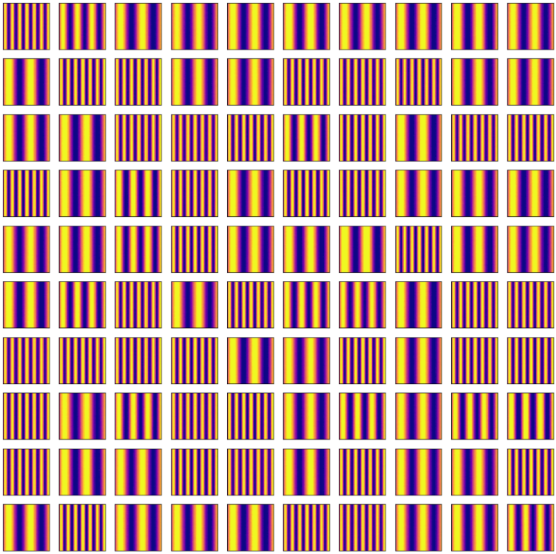}
    \caption{Proposed processing temperatures generated under conditions of test data}
    \label{fig:testdata_condition_tempr}
\end{figure}

\begin{figure}[tb]
    \centering
    \includegraphics[width=0.75\columnwidth]{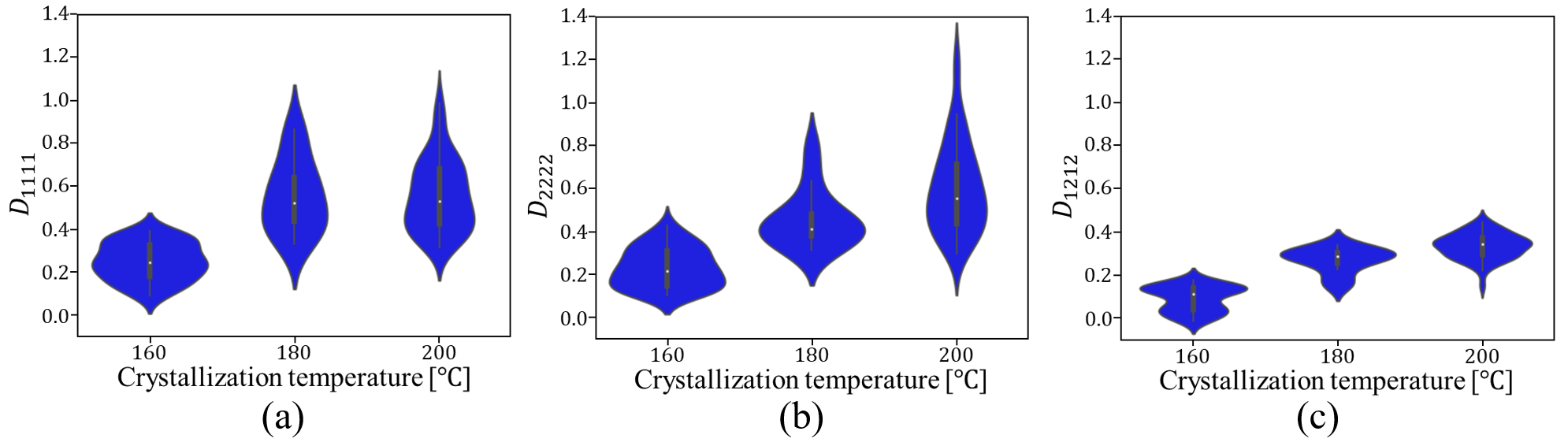}
    \caption{Relationship between crystallization temperature and elasticity matrix (a) $D_{1111}$, (b) $D_{2222}$ and (c) $D_{1212}$ generated under conditions of test data}
    \label{fig:generatedbytest_violin}
\end{figure}

\begin{figure}[tb]
    \centering
    \includegraphics[width=0.4\columnwidth]{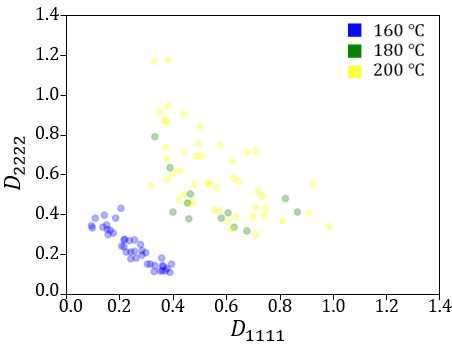}
    \caption{Relationship between $D_{1111}$ and $D_{2222}$ generated unedr conditions of test data for each crystallization temperature}
    \label{fig:testsave_dtemp}
\end{figure}

\begin{figure}[tb]
    \centering
    \includegraphics[width=0.75\columnwidth]{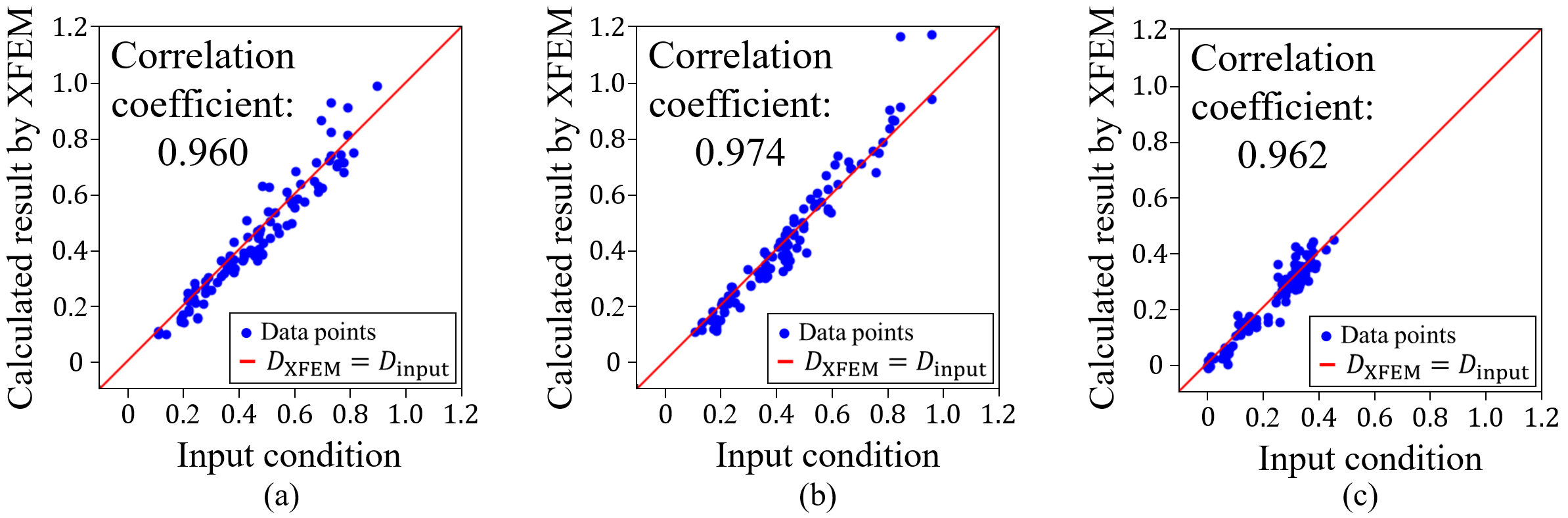}
    \caption{Correlation coefficients of elasticity matrix (a) $D_{1111}$, (b) $D_{2222}$ and (c) $D_{1212}$ generated uneder conditions of test data}
    \label{fig:generatedbytest_correlation}
\end{figure}

Firstly, we confirmed the trained model by using the condition of the test data. 
Figs. \ref{fig:testdata_condition}--\ref{fig:generatedbytest_correlation} show the results.
Fig. \ref{fig:testdata_condition} shows the generated images of microstructures and Fig. \ref{fig:testdata_condition_tempr} shows the generated images that indicate the processing temperature. 
Fig. \ref{fig:generatedbytest_violin} shows the relationship between the proposed temperature and the elasticity matrix $\boldsymbol{D}$ obtained by homogenization analysis using XFEM again for the generated microstructure. 
Fig. \ref{fig:testsave_dtemp} is a scatter plot with $D_{1111}$ on the horizontal axis and $D_{2222}$ on the vertical axis, plotted in different colors for different crystallization temperature. 
From Figs.  \ref{fig:generatedbytest_violin} and \ref{fig:testsave_dtemp}, the relationship between crystallization temperature and elasticity matrix $\boldsymbol{D}$ is generally appropriate because the tendencies of the training data and test data are similar. 
Fig. \ref{fig:generatedbytest_correlation} compares the values of each component of the elasticity matrix $\boldsymbol{D}$ obtained by homogenization analysis using XFEM of the generated microstructure again with the condition given for generation. 
The red line indicates the ideal situation that the elasticity matrix $\boldsymbol{D}$ obtained by homogenization analysis using XFEM for the generated microstructure is equal to the elasticity matrix $\boldsymbol{D}$ given as the condition, i.e., when $D_\text {XFEM} = D_\text {input}$.
Where $D_\text {XFEM}$ is the element of the elasticity matrix $\boldsymbol{D}$ obtained by homogenization analysis using XFEM of the generated microstructure again, $D_\text {input}$ is the element of the elasticity matrix $\boldsymbol{D}$ given for generation.
From the correlation coefficients and each plot, the elasticity matrix $\boldsymbol{D}$ obtained by the homogenization analysis using XFEM shows good agreement with the condition given for generation. 
The correlation coefficients indicates a strong positive correlation, which is considered that the predicted microstructure is appropriate for the given condition.

\clearpage

\subsection{Results for conditions not included in the dataset}
\begin{figure}[h]
    \centering
    \includegraphics[width=0.8\columnwidth]{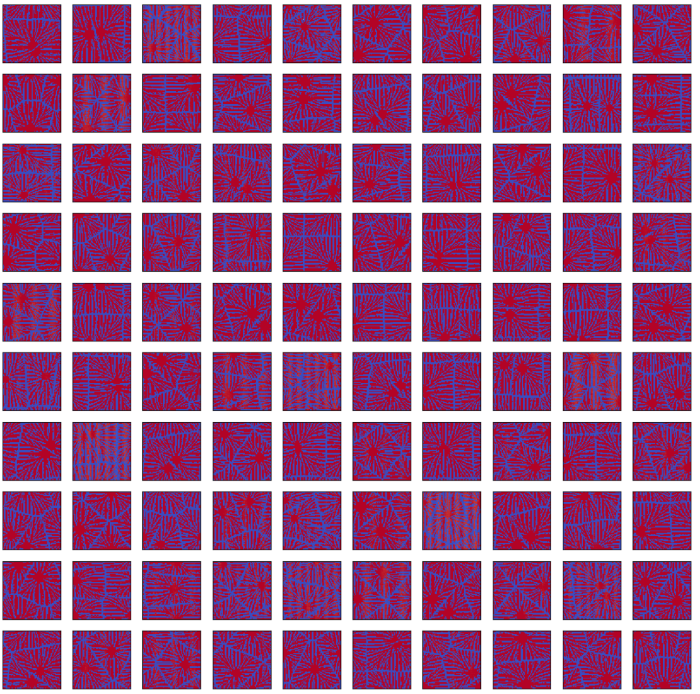}
    \caption{Predicted microstructures generated under the conditions that are not included in the dataset}
    \label{fig:midtestdata_condition}
\end{figure}

\begin{figure}[tb]
    \centering
    \includegraphics[width=0.8\columnwidth]{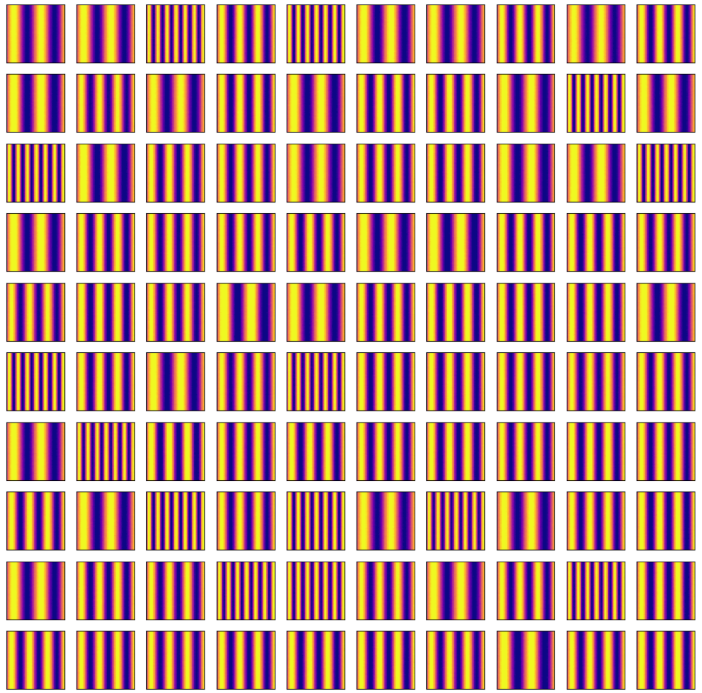}
    \caption{Proposed processing temperatures generated under the conditions that are not included in the dataset}
    \label{fig:midtestdata_condition_tempr}
\end{figure}

\begin{figure}[tb]
    \centering
    \includegraphics[width=0.75\columnwidth]{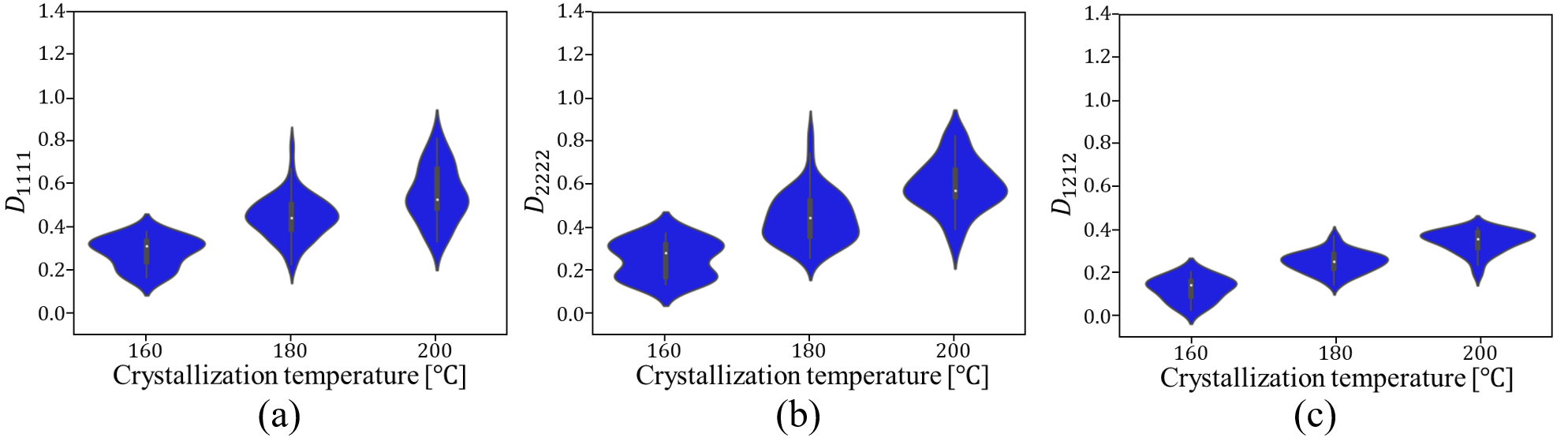}
    \caption{Relationship between crystallization temperature and elasticity matrix (a) $D_{1111}$, (b) $D_{2222}$ and (c) $D_{1212}$ generated under the conditions that are not included in the dataset}
    \label{fig:generatedbymidtest_violin}
\end{figure}

\begin{figure}[tb]
    \centering
    \includegraphics[width=0.4\columnwidth]{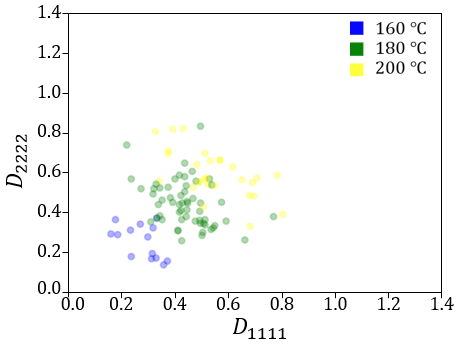}
    \caption{Relationship between $D_{1111}$ and $D_{2222}$ generated under the conditions that are not included in the dataset for each crystallization temperature}
    \label{fig:intestsave_dtemp}
\end{figure}

\begin{figure}[tb]
    \centering
    \includegraphics[width=0.9\columnwidth]{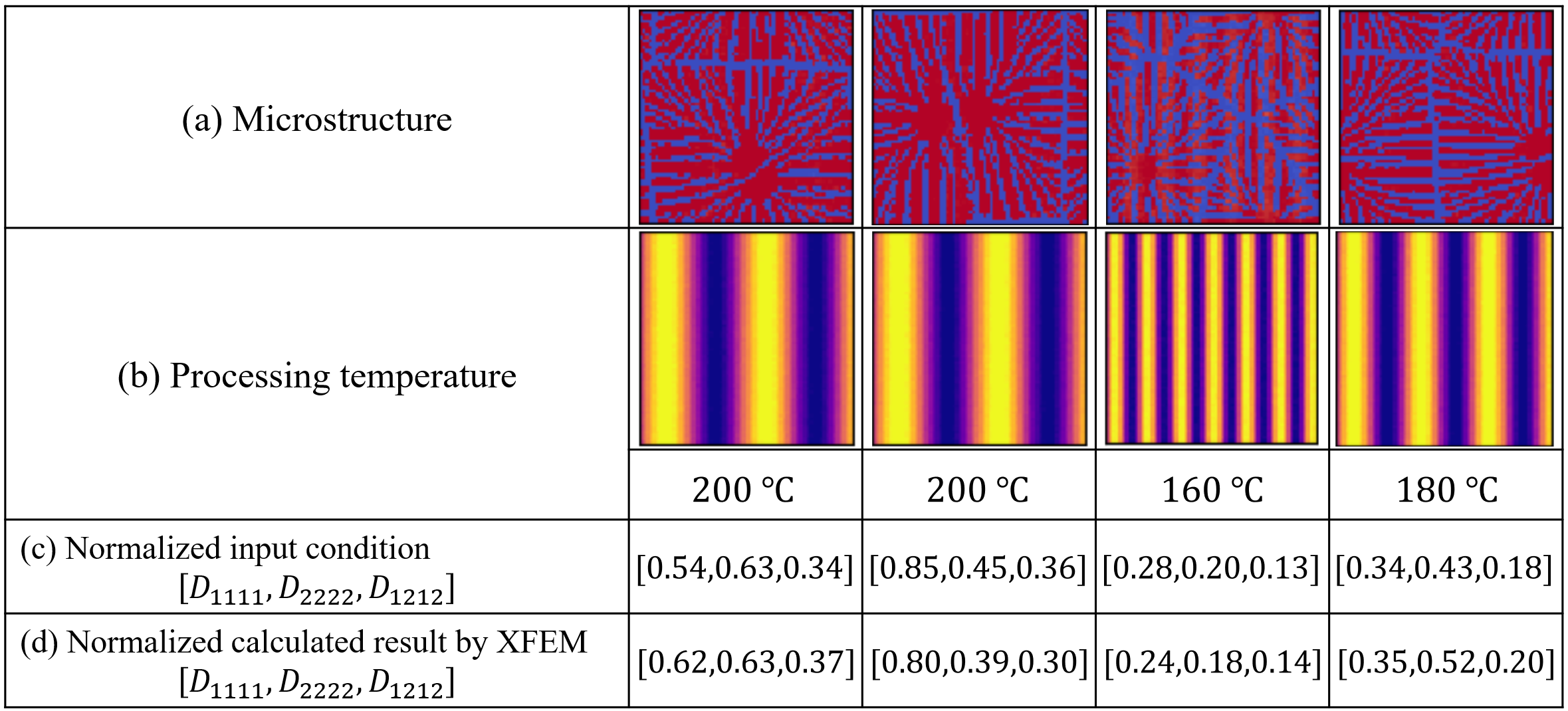}
    \caption{Results of generation under the conditions that are not included in the dataset}
    \label{fig:intestsamplingtemp}
\end{figure}

\begin{figure}[bt]
    \centering
    \includegraphics[width=0.75\columnwidth]{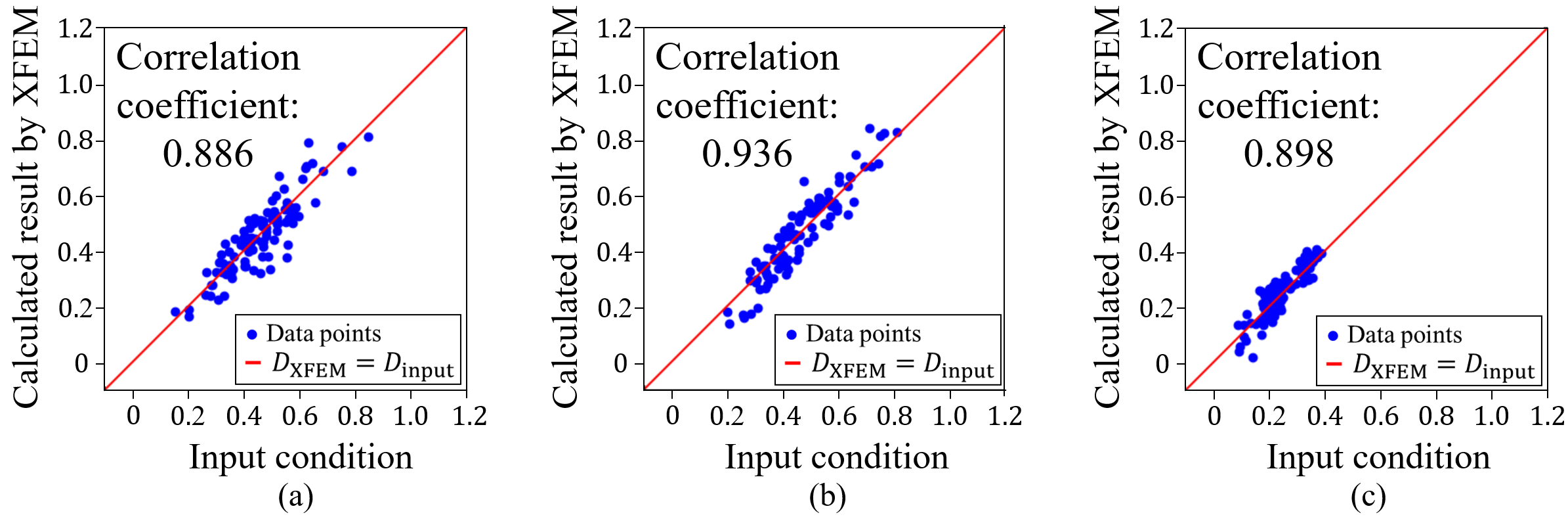}
    \caption{Correlation coefficients of the elasticity matrix (a) $D_{1111}$, (b) $D_{2222}$ and (c) $D_{1212}$ generated under the conditions that are not included in the dataset}
    \label{fig:generatedbymidtest_correlation}
\end{figure}

We confirm the results for the conditions not included in the dataset.
The results are shown in Figs. \ref{fig:midtestdata_condition}--\ref{fig:generatedbymidtest_correlation}. 
Fig. \ref{fig:midtestdata_condition} shows the generated images of microstructures and Fig. \ref{fig:midtestdata_condition_tempr} shows the generated images that indicate the processing temperature. 
Fig. \ref{fig:generatedbymidtest_violin} shows the relationship between the proposed temperature and the elasticity matrix $\boldsymbol{D}$ obtained by homogenization analysis using XFEM for the generated microstructure.
Fig. \ref{fig:intestsave_dtemp} is a scatter plot with $D_{1111}$ on the horizontal axis and $D_{2222}$ on the vertical axis, plotted in different colors for different crystallization temperatures.
Figs. \ref{fig:generatedbymidtest_violin} and \ref{fig:intestsave_dtemp} show that the relationship between crystallization temperature and the elasticity matrix $\boldsymbol{D}$ is generally similar to that of the training and test data, even when the data are generated using the conditions not included in the dataset. 
Fig. \ref{fig:intestsamplingtemp} shows examples of the generated results. 
The model developed outputs images of microstructures and IPPTs. 
Fig. \ref{fig:intestsamplingtemp} shows (a) the generated microstructures, (b) IPPTs, (c) the conditions and (d) the elasticity matrix $\boldsymbol{D}$ values obtained by homogenization analysis using XFEM for the generated microstructure.
As shown in Fig. \ref{fig:intestsamplingtemp}(a), the conditional diffusion model can represent the detailed microstructure of dendrite crystals of thermoplastic resins.
By homogenization analysis with XFEM to obtain the elasticity matrix $\boldsymbol{D}$ for the predicted microstructure, we find that elasticity matrix $\boldsymbol{D}$ is close to the condition given for generation.
Fig. \ref{fig:generatedbymidtest_correlation} shows the elasticity matrix $\boldsymbol{D}$ for each component.
The horizontal axis represents the elasticity matrix $\boldsymbol{D}$ obtained by homogenization analysis using XFEM for the generated microstructure, and the vertical axis is the condition given for generation. 
The correlation coefficients and each plot indicate that the values of elasticity matrix $\boldsymbol{D}$ obtained by homogenization analysis using XFEM agree with the condition given for generation with good accuracy. 
The correlation coefficients imply a strong positive correlation between input condition and calculated result by XFEM, suggesting appropriate microstructure for the given condition.

\clearpage
\subsection{Demonstration of the proposed conditional diffusion model}
\begin{figure}[h]
    \centering
    \includegraphics[width=0.85\columnwidth]{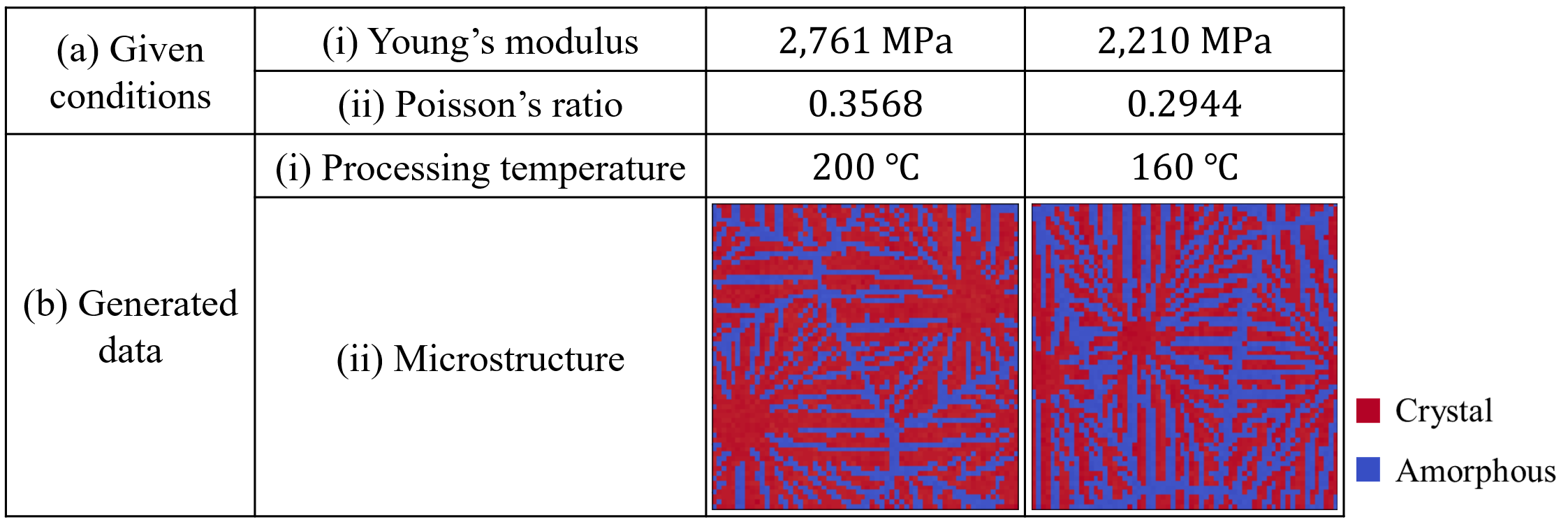}
    \caption{Proposed processing temperatures and predicted microstructures obtained using conditional diffusion model}
    \label{fig:table_example}
\end{figure}

The conditional diffusion model developed in this study can propose the optimal processing temperature and predict the microstructure for given Young's modulus and Poisson's ratio.
The advantage of this model is that it outputs not only the processing temperature but also the microstructure as images.
Fig. \ref{fig:table_example} shows example of the model demonstration.
Fig. \ref{fig:table_example} shows (a)(i) the specified values of Young's modulus, (a)(ii) the specified values of Poisson's ratio, (b)(i) the proposed temperatures and (b)(ii) the predicted microstructures of thermoplastic resins.
We train the model with the conditions obtained by homogenization analysis using XFEM on $320 {\ \mathrm{pixels}}\times320 {\ \mathrm{pixels}}$ before image compression.
Table \ref{trainingdata320} shows the number and size of data used for training.
The relationship between Young's modulus and Poisson's ratio and the calculation of the elasticity matrix $\boldsymbol{D}$ is described in \ref{secA1}.
As shown in Fig. \ref{fig:table_example}, 200 ${}^\circ$C is proposed for a large Young's modulus of $2,761 \,\text {MPa}$, whereas 160 ${}^\circ$C is suggested for a small Young's modulus of $2,210 \,\text {MPa}$.
Thus, the results show that the temperature tends to increase with Young's modulus.

\begin{table}[h]
    \centering
    \caption{
       {Data used to train \text{$320{\ \mathrm{pixels}}\times320{\ \mathrm{pixels}}$}}{\textmd{}}
    }
    \label{trainingdata320}
    \begin{tabular}{c c}\hline
        Number of training data & 13,908\\
        Number of validation data & 1,728\\
        Number of test data & 432\\
        Image size  &  $64{\ \mathrm{pixels}}\times64{\ \mathrm{pixels}}$\\
        Minibatch size & 5 \\ 
        Number of epoch & 549 \\ \hline
    \end{tabular}
\end{table}

\clearpage

\section{Conclusions}
\label{sec4}
In this study, we developed a conditional diffusion model that can propose the optimal processing temperature and predict the microstructure for the desired elastic constants of thermoplastic resins.
As a result of using the developed model, the following three were confirmed:
\begin{itemize}
  \item By training the pattern that indicates the processing temperature together with the images of the microstructure, the model could predict not only the processing temperature but also the microstructure when the elastic constants were given. 
  \item The developed model proposed high temperatures with high Young's modulus. At high temperatures, the predicted crystal structures were obtained with thicker crystal chains.
  \item Even when the conditions were not included in the dataset, complicated dendritic patterns whose elastic constants were reasonable were reproduced.
\end{itemize}
In conclusion, the conditional diffusion model developed in this study can propose the optimal processing temperature and predict the microstructure of thermoplastic resins that satisfies the desired mechanical properties. 
This study enables us to realize an inverse design that proposes the process parameters related to how the material should be made so as to satisfy the mechanical properties.

This model can be applied to other materials, process parameters and mechanical properties by replacing the data used for training, such as the microstructure, processing temperature and elastic constants.

\section*{Acknowledgements}
This study was partially supported by the Japan Society for the Promotion of Science (JSPS) KAKENHI (Grant Number 22K14489).

\section*{Conflict of interest}
The authors declare that they have no conflicts of interest to disclose.

\section*{Contributions}
A.I. conducted Data curation, Formal analysis, Investigation, Methodology, Software, Visualization, Validation, Writing - Original Draft, Writing - Review and Editing.
R.H. conducted Conceptualization, Funding acquisition, Investigation, Methodology.
T.Y. conducted Conceptualization, Funding acquisition, Methodology.
K.E. conducted Conceptualization, Investigation, Methodology, Software, Validation.
Y.K. conducted Investigation, Methodology, Software, Validation, Writing - Review and Editing.
M.S. conducted Investigation, Methodology, Software, Validation.
M.M. conducted Conceptualization, Funding acquisition, Project administration, Resources, Writing - Review and Editing, Supervision.

\section*{Appendix}

\subsection*{Homogenization analysis and elasticity matrix}\label{secA1}
The input data for homogenization analysis using XFEM are  phase-field variables, grid information, material properties and boundary conditions. 
The relationship between stress and strain is expressed as
\begin{equation}
\label{sigdep}
\left\{\begin{array}{c}
\sigma_{11} \\
\sigma_{22} \\
\tau_{12} 
\end{array}\right\}=\left[\begin{array}{cccccc}
D_{1111} & D_{1122} & D_{1112} \\
D_{1122} & D_{2222} & D_{2212} \\
D_{1112} & D_{2212} & D_{1212} \\
\end{array}\right]\left\{\begin{array}{c}
\varepsilon_{11} \\
\varepsilon_{22} \\
\gamma_{12} 
\end{array}\right\},
\end{equation}
where $\sigma_{11}$ and $\sigma_{22}$ are the normal stress, $\tau_{12}$ is the shear stress, $D_{1111}$, $D_{1122}$, $D_{1112}$, $D_{2222}$, $D_{2212}$ and $D_{1212}$ are the components of  the elasticity matrix $\boldsymbol{D}$, $\varepsilon_{11}$ and $\varepsilon_{22}$ are the normal strain and $\gamma_{12}$ is the shear strain. 
For each component of shear strain, the engineering shear strain $\gamma_{12}$ , which is twice the amount of the shear component of the strain tensor, is used.
By homogenization analysis using XFEM, we obtain $D_{1111}$, $D_{2222}$ and $D_{1212}$ by applying unit strain in the following three patterns:
\begin{subequations}
\begin{equation}
\label{sig11}
\left\{\begin{array}{c}
\varepsilon_{11} \\
\varepsilon_{22} \\
\gamma_{12} 
\end{array}\right\}=\left\{\begin{array}{c}
1 \\
0 \\
0 
\end{array}\right\},
\end{equation}
\begin{equation}
\label{sig22}
\left\{\begin{array}{c}
\varepsilon_{11} \\
\varepsilon_{22} \\
\gamma_{12} 
\end{array}\right\}=\left\{\begin{array}{c}
0 \\
1 \\
0 
\end{array}\right\},
\end{equation}
\text and
\begin{equation}
\label{tau12}
\left\{\begin{array}{c}
\varepsilon_{11} \\
\varepsilon_{22} \\
\gamma_{12} 
\end{array}\right\}=\left\{\begin{array}{c}
0 \\
0 \\
1 
\end{array}\right\}.
\end{equation}
\end{subequations}
The mean stress distribution in each direction is written as macroscopic stress in the out-file output of homogenization analysis using XFEM.
Since we provide unit strains, $D_{1111}$ is $\sigma_{11}$ when Eq. (\ref{sig11}) is substituted into Eq. (\ref{sigdep}), $D_{2222}$ is $\sigma_{22}$ when Eq. (\ref{sig22}) is substituted into Eq. (\ref{sigdep}) and $D_{1212}$ is $\tau_{12}$ when Eq. (\ref{tau12}) is substituted into Eq. (\ref{sigdep}).

Owing to the assumption of isothermal forming as the simplest process condition, Young's modulus and Poisson's ratio are expressed using the mean values of $D_{1111}$ and $D_{2222}$:
\begin{equation}
    E = \frac{D_{1212}\left(\frac{3(D_{1111}+D_{2222})}{2}-4D_{1212}\right)}{\frac{D_{1111}+D_{2222}}{2}-D_{1212}},
\end{equation}
\begin{equation}
    \nu = \frac{\frac{D_{1111}+D_{2222}}{2}-2D_{1212}}{2\left(\frac{D_{1111}+D_{2222}}{2}-D_{1212}\right)}.
\end{equation}

\bibliography{sample}

\end{document}